\documentclass[%
 reprint,
superscriptaddress,
showpacs,preprintnumbers,
nofootinbib,
 amsmath,amssymb,
 prl,
]{revtex4-2}

\usepackage{graphicx}
\usepackage{dcolumn}
\usepackage{bm}
\usepackage{epstopdf}
\usepackage{epsfig}
\usepackage{xcolor}
\usepackage{hyperref}
\hypersetup{%
	colorlinks=true, citecolor=blue, filecolor=blue, linkcolor=blue, urlcolor=blue}

\begin{document}

\preprint{This draft was prepared using the LaTeX style file belonging to the APS}

\title{Universal scaling of nanoparticle deposition by colloidal droplet drying}

\author{Feifei Qin}
\affiliation{Institute of Extreme Mechanics and School of Aeronautics, Northwestern Polytechnical University, Xi’an 710072, PR China}
\affiliation{Chair of Building Physics, Department of Mechanical and Process Engineering, ETH Zürich (Swiss Federal Institute of Technology in Zürich), Zürich 8092, Switzerland}

\author{Linlin Fei}
\email{linfei@ethz.ch}
\affiliation{Chair of Building Physics, Department of Mechanical and Process Engineering, ETH Zürich (Swiss Federal Institute of Technology in Zürich), Zürich 8092, Switzerland}

\author{Jianlin Zhao}
\affiliation{College of Petroleum Engineering, China University of Petroleum in Beijing, PR China}
\affiliation{Chair of Building Physics, Department of Mechanical and Process Engineering, ETH Zürich (Swiss Federal Institute of Technology in Zürich), Zürich 8092, Switzerland}

\author{Qinjun Kang}
\affiliation{Earth and Environment Sciences Division (EES-16), Los Alamos National Laboratory (LANL), Los Alamos, NM 87545, USA}

\author{Sauro Succi}
\affiliation{Center for Life Nano Science at La Sapienza, Fondazione Istituto Italiano di Tecnologia, Viale Regina Margherita 295, 00161, Roma, Italy}
\affiliation{Physics Department and Institute for Applied Computational Science, John A. Paulson School of Applied Science and Engineering, Harvard University, Oxford Street 29, Cambridge, MA 02138, USA}

\author{Dominique Derome}
\affiliation{Dep. of Civil and Building Engineering, Université de Sherbrooke, Sherbrooke Qc$^\circ $ J1K 2R1 Canada}

\author{Jan Carmeliet}
\affiliation{Chair of Building Physics, Department of Mechanical and Process Engineering, ETH Zürich (Swiss Federal Institute of Technology in Zürich), Zürich 8092, Switzerland}
\date{\today}

\begin{abstract}
  We present a comprehensive study of nanoparticle deposition from drying of colloidal droplets. By means of lattice Boltzmann modeling and theoretical analysis, various deposition patterns, including mountain-like, uniform and coffee ring, as well as un-/symmetrical multiring/mountain-like patterns are achieved. The ratio of nanoparticles deposited at droplet peripheries and center is proposed to quantify different patterns. Its value is controlled by the competition between the capillary flow and nanoparticle diffusion, leading to a linear dependence on an effective Péclet number, across over three orders of magnitude. Remarkably, the final deposition pattern can be predicted based on material properties only.
\end{abstract}                             
\maketitle

\begin{figure}[!htbp]
	\center {
		{\epsfig{file=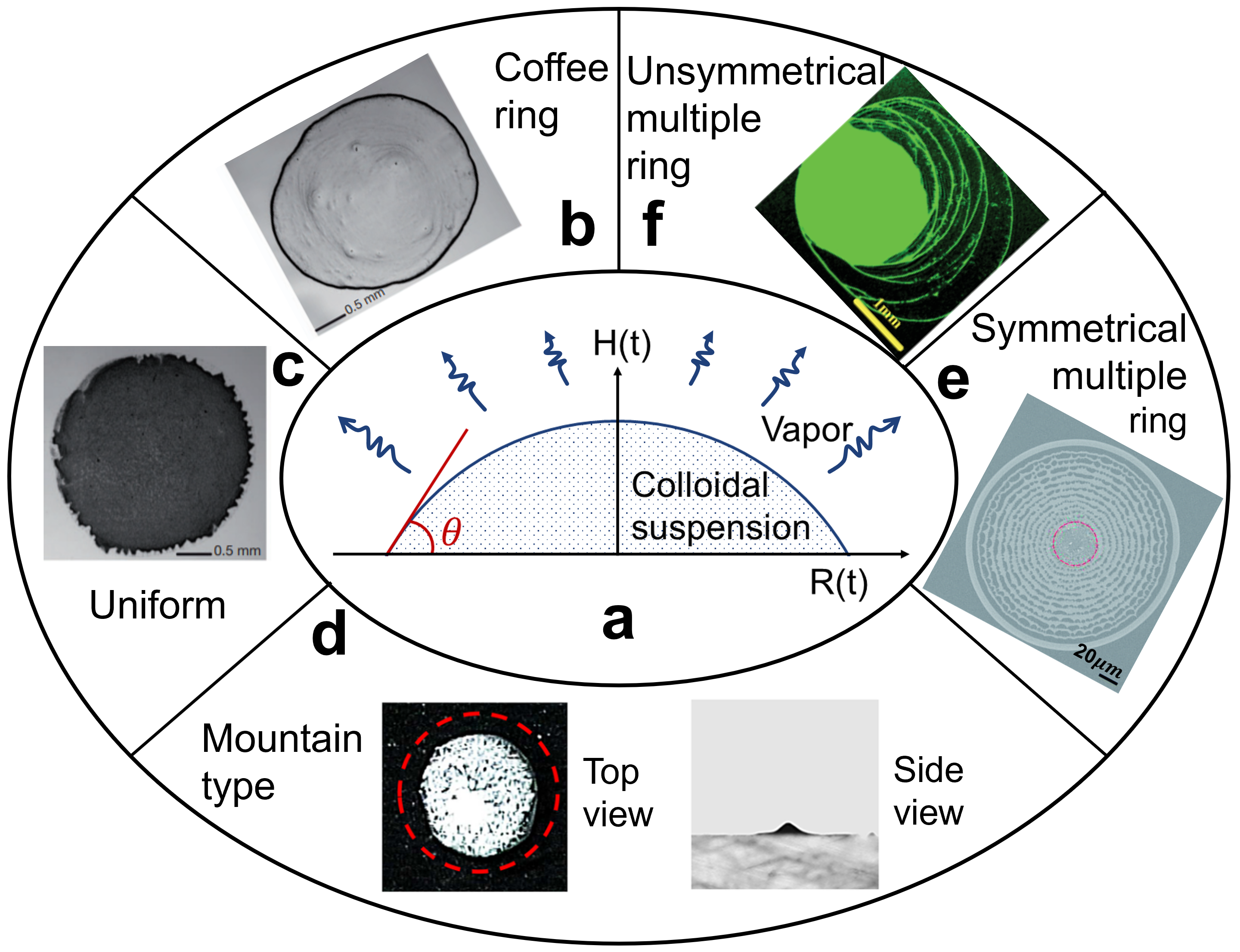,width=0.4\textwidth,clip=}}\hspace{0.0cm}  
	}
	\caption{Illustration of various nanoparticle deposition patterns after drying of a colloidal droplet. (a) Colloidal droplet drying on a flat surface. (b) Coffee ring \cite{Yunker2011}. (c) Uniform deposition \cite{Yunker2011}. (d) Mountain type \cite{Li2014c}. (e) Symmetrical multiple rings \cite{Yang2014}. (f) Unsymmetrical multiple rings \cite{Maheshwari2008}.}
	\label{Fig01}
\end{figure}

Evaporation of colloidal suspension with induced nanoparticle deposition (Fig. \ref{Fig01}) is ubiquitously observed in nature and widely applied in the industry \cite{Bonn2009,Mampallil2018}. The best-known example is the coffee ring formation (Fig. \ref{Fig01}b), where most of particles are swept to the pinned contact line of the droplet. The physical mechanism first explained by Deegan et al. \cite{Deegan1997} is that, an internal capillary flow induced by different evaporation rates over the droplet surface drives the nanoparticles from the apex to the periphery of the droplet. Multiring patterns (Fig. \ref{Fig01}e-f) are reported when stick-slip behavior occurs caused by sufficient contact angle hysteresis \cite{Shmuylovich2002,Maheshwari2008,Moffat2009,Yang2014}. Opposite to ring depositions, a mountain-like deposition (Fig. \ref{Fig01}d) is observed when a droplet evaporates with a receding contact line \cite{Willmer2010,Li2014c}. Uniform depositions (Fig. \ref{Fig01}c) can be achieved by using non-spherical nanoparticles \cite{Yunker2011} or adjusting the pH value of the liquid \cite{Bhardwaj2010}. Other deposition patterns, such as volcano-like and hexagonal cells type have also been reported when drying a polymer solution \cite{Kajiya2009} or surfactant-laden aqueous drop \cite{Truskett2003}. 

To understand the underlying mechanisms of diverse depositions, Deegan et al. \cite{Deegan2000a} first applied a theoretical model to predict liquid flow and particle distribution in evaporating droplets. Hu \& Larson \cite{Hu2002,Hu2005b} proposed a simple approximation of the evaporation rate and also investigated the microfluid flow based on theoretical and finite element analyses. The effects of Marangoni stress on microfluid flow, evaporation rate and deposition pattern were further elaborated in \cite{Hu2005a,Bhardwaj2009a}. Frastia et al. \cite{Frastia2011} proposed a dynamic model to predict the multiring deposition considering varying particle concentration and evaporation rate. Man \& Doi \cite{Man2016} predicted a ring to mountain-like transition based on the modeling of fluid flow and contact line motion, which is further extended to multiple ring deposition \cite{Wu2018}. Kaplan \& Mahadevan \cite{Kaplan2014} developed a multiphase model by coupling the inhomogeneous evaporation rate, droplet internal flow, and local nanoparticle concentration. They also proposed a dimensionless number to characterize the transition from ring to uniform deposition. The deposition dynamics is found to depend on many parameters, such as surface heterogeneity, contact angle, surface tension, liquid viscosity and nanoparticle diffusivity. 

Despite the extensive efforts of previous studies, a comprehensive analysis of the effects of various surface/liquid properties is lacking, as well as the construction of a universal governing law quantitatively characterizing the different deposition patterns. In this Letter we accomplish precisely this task; by combining colloidal particle modeling \cite{Zhang2021a,Zhao2018b,Qin2019a,Nath2021} with a lattice Boltzmann treatment \cite{Li2016d,Huang2021,Qin2019,Wohrwag2018,Succi2018,Benzi2009,Higuera1989a} of the fluid droplet, we unveil a universal relation linking the characteristic features of the deposition pattern to a single dimensionless parameter.

The proposed double-distribution multiple-relaxation-time lattice Boltzmann model (MRT LBM) for two-phase flow and nanoparticle modeling is introduced in Supplementary Material with two validation cases, i.e., a 1D colloidal liquid drying (Fig. S1, S2) and a 2D colloidal droplet drying on flat surface with single stick-slip process (Fig. S3-S6) \cite{Yunker2011,Stauber2015a}. Besides the methodological aspects, we elaborate on the physical mechanisms of various drying and deposition patterns and transition criteria among each, as summarized in Fig. \ref{Fig01}. Fig. \ref{Fig01}a illustrates the colloidal droplet drying on a flat surface, while Fig. \ref{Fig01}b-f show the potential deposition patterns ranging from coffee ring, uniform, mountain type to symmetrical and unsymmetrical multiple rings. 

\begin{figure}[!htbp]
	\center {
		{\epsfig{file=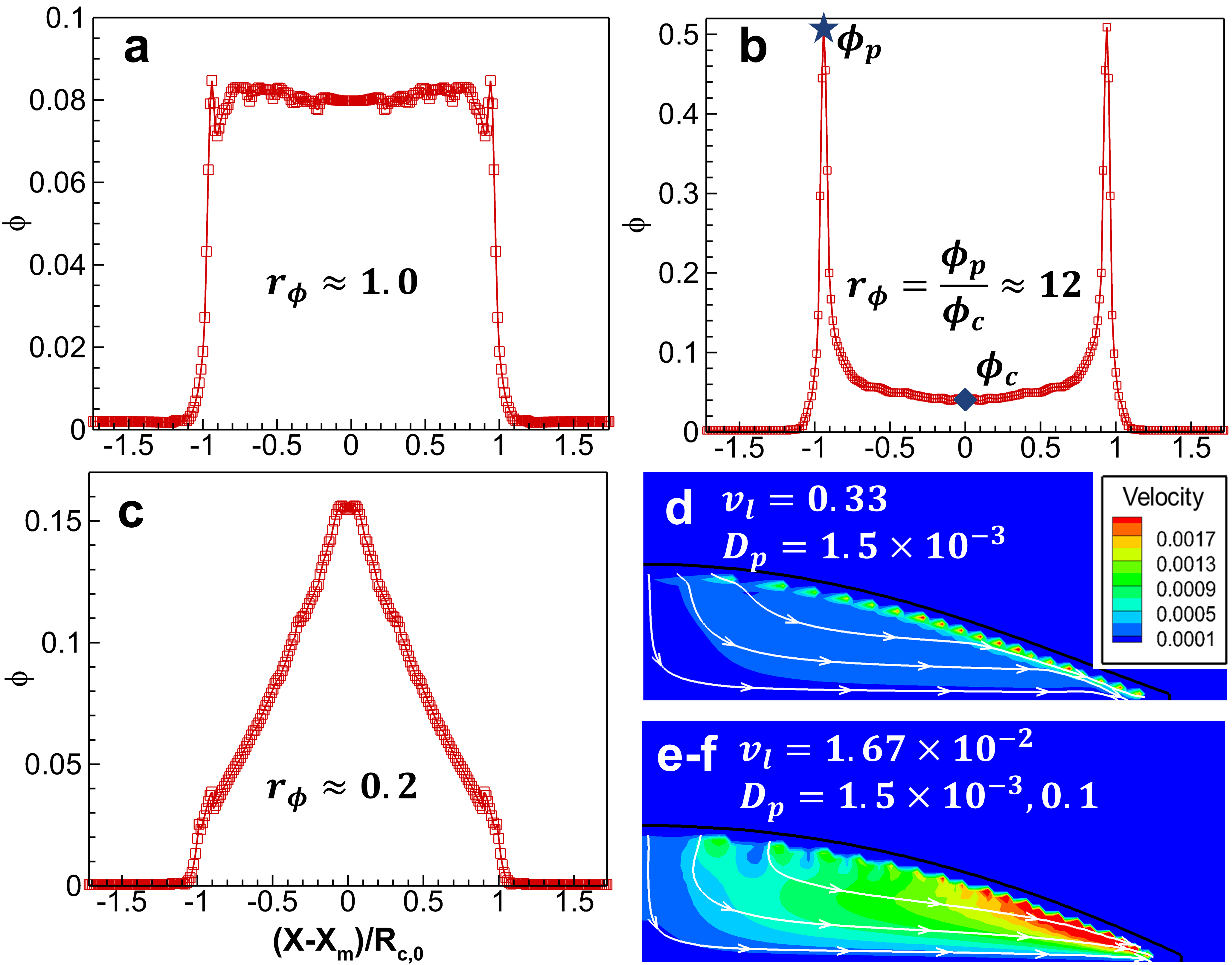,width=0.48\textwidth,clip=}}\hspace{0.0cm}  
	}
	\caption{ Different deposition patterns with corresponding droplet internal flows. (a-c) Uniform, ring, and mountain-like depositions (SM01-03). (d-f) Streamlines and velocity contour inside the droplet corresponding to (a-c) with given liquid kinematic viscosity ${v_l}$ and nanoparticle diffusion coefficient ${D_p}$. The initial concentration of all 3 cases is ${\phi _0} = 0.5\% $, the surface tension is $\sigma  = 1.05 \times {10^{ - 2}}$, the initial and receding contact angles are ${\theta _{eq}} = 35^\circ $ and ${\theta _R} = 5^\circ $. All variables are given in lattice units. }
	\label{Fig02}
\end{figure}

We first focus on the simple deposition patterns corresponding to Fig. \ref{Fig01}b-d with the initial colloidal nanoparticle concentration ${\phi _0} = 0.5\% $, surface tension $\sigma  = 1.05 \times {10^{ - 2}}$, the equilibrium and receding contact angles ${\theta _{eq}} = 35^\circ $ and ${\theta _R} = 5^\circ $. To quantify the various deposition, we propose to use the deposition ratio ${r_\phi } = {\phi _p}/{\phi _c}$ \cite{Yunker2011}, where ${\phi _p}$ and ${\phi _c}$ are the depositions at the droplet periphery and center, respectively.  The horizontal axis of deposition is normalized by droplet initial contact radius ${R_{c,0}}$ respect to the center location ${X_m}$ , i.e., $(X - {X_m})/{R_{c,0}}$. As shown in Fig. \ref{Fig02}a-c, ${r_\phi } \approx 1$ corresponds to a uniform deposition (Supplementary movie SM01), while ${r_\phi } \gg 1$ and ${r_\phi } \ll 1$ represent coffee ring (SM02) and mountain-like (SM03) deposition patterns, respectively. In the isothermal evaporation, two major nanoparticle transport processes take place during droplet drying, i.e., capillary transport and diffusion. The former is induced by an unequal drying rate at the droplet surface (Fig. S6) and depends on liquid properties and drying conditions, while the latter is induced by concentration differences and depends on the diffusion coefficient determined by Stokes–Einstein equation \cite{Costigliola2019,Qin2019}. The internal capillary flow for different liquid/nanoparticle parameters corresponding to deposition patterns in Fig. \ref{Fig02}a-c is shown in Fig. \ref{Fig02}d-f. Comparison between Fig. \ref{Fig02}d and e indicates that strong capillary flow at low liquid viscosity leads to more deposition at the contact line. Fig. \ref{Fig02}e and f show an identical capillary flow owing to the same fluid properties. At a higher diffusion coefficient (Fig. \ref{Fig02}f), the diffusion transport is more significant so that nanoparticles are effectively redistributed to be uniform in the stick period before their deposition, and mountain-like deposition (Fig. \ref{Fig02}c) finally forms due to the receding of the contact line during the slip phase, bringing nanoparticles to the center (SM03).

\begin{figure}[!htbp]
	\center {
		{\epsfig{file=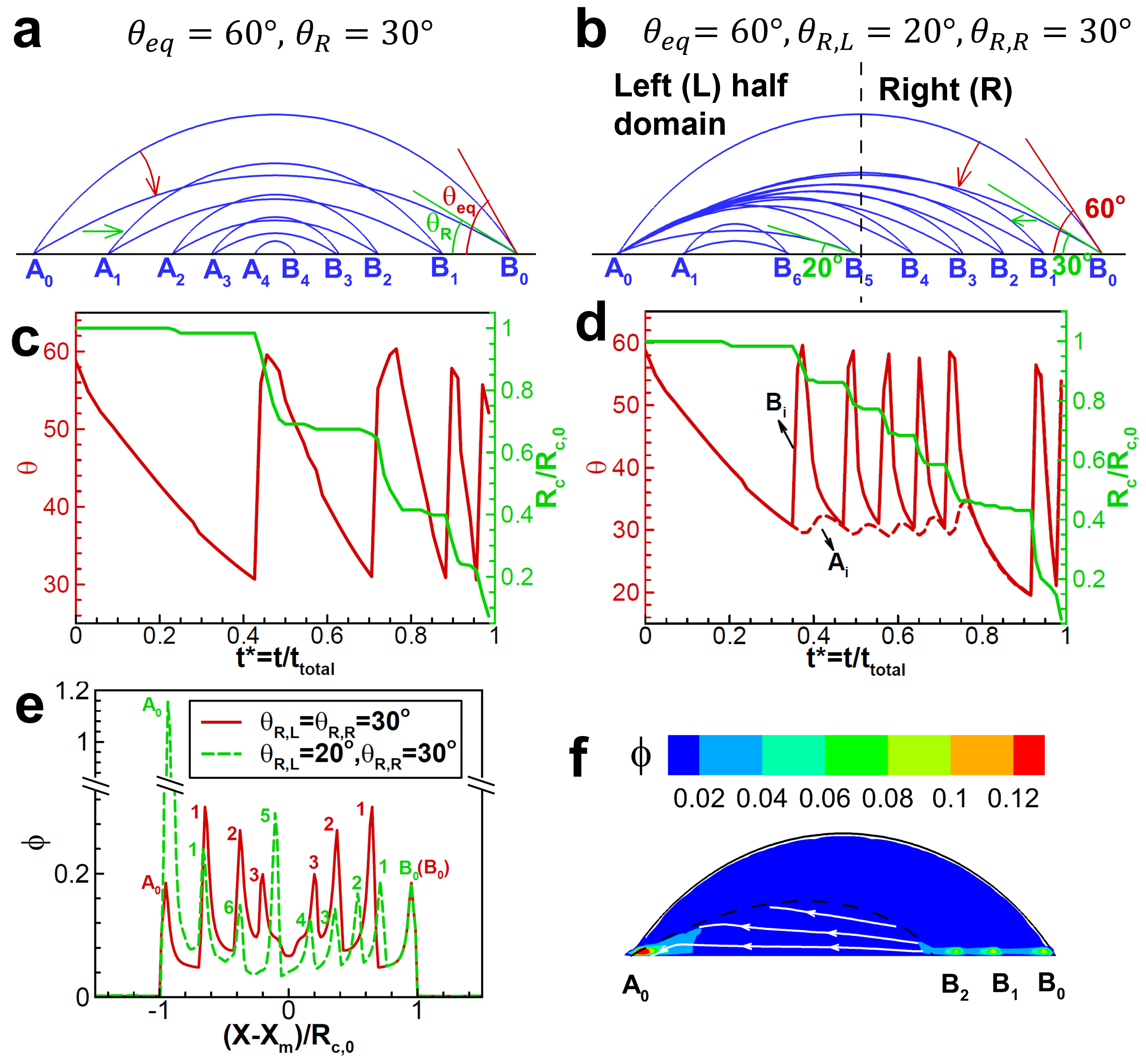,width=0.48\textwidth,clip=}}\hspace{0.0cm} \\ 
	}
	\caption{Multiring formation considering spatially non-/uniform contact angle hysteresis. (a-b) Consecutive stick-slip behavior during colloidal droplet drying with uniform contact angle hysteresis range ${\theta _{eq}} = 60^\circ ,{\theta _R} = 30^\circ $ and nonuniform ones, i.e. ${\theta _{eq}} = 60^\circ $ with ${\theta _{R,L}} = 20^\circ $ and ${\theta _{R,R}} = 30^\circ $ at the left and right half domains (SM04-05). (c-d) Evolution of contact angle $\theta $ and normalized contact radius ${R_c}/{R_{c,0}}$ during drying corresponding to cases (a-b). (e) Comparison of final deposition patterns in cases (a-b). (f) Nanoparticle concentration contour at two time frames during drying in case b, with streamlines showing the capillary transport contributing to the nanoparticle accumulation at contact point ${A_0}$.}
	\label{Fig03}
\end{figure}

Further, to achieve the complex un-/concentric multiple ring depositions as shown in Fig. \ref{Fig01}e-f, we model the drying process of a colloidal droplet with consecutive stick-processes. The local nanoparticle deposition is assumed to have negligible influence on the receding contact angle. Compared with the single stick-slip motion discussed above, the governing rules of contact angle for the multiple stick-slip processes are \cite{Man2016}:

\begin{equation}\label{eq:theta_t}
\theta (t) = \left\{ {\begin{array}{*{20}{c}}
		{{\theta _m},{\rm{   }}pinning{\rm{      }} \Leftarrow {\rm{    }}{\theta _m} \ge {\theta _{eq}}{\rm{    }}}\\
		{{\theta _{eq}},{\rm{ }}receding{\rm{     }} \Leftarrow {\rm{    }}{\theta _m} \le {\theta _R}{\rm{    }}}
\end{array}} \right.
\end{equation}
where ${\theta _m}$ is the measured local contact angle at the liquid-gas interface. We first consider “case a” where the hysteresis range (${\theta _{eq}} = 60^\circ $, ${\theta _R} = 30^\circ $) is uniform over the entire simulation domain. As shown in Fig. \ref{Fig03}a, the droplet follows the stick-slip cycles and the location of two contact points ${A_i}$ and ${B_i}$ at both sides of the droplet remain symmetric during drying. Fig. \ref{Fig03}c shows the contact radius stays unchanged when the contact angle is decreasing from ${\theta _{eq}}$ to ${\theta _R}$ during stick phase, while it drops during the contact line slip with increasing contact angle. The pinning time gradually decreases within each stick-slip process due to decreasing droplet volume. The final deposition is shown in Fig. \ref{Fig03}e with solid red lines (SM04). 

When the surface roughness is non-uniform, a variation of the contact angle hysteresis range occurs in space \cite{Maheshwari2008,Moffat2009}. As an example, we consider “case b” where the left- and right-half domains show the same initial contact angle ${\theta _{eq}} = 60^\circ $ but with different receding contact angles ${\theta _{R,L}} = 20^\circ $ and ${\theta _{R,R}} = 30^\circ $. Fig. \ref{Fig03}b shows the drying process is asymmetric initially. After contact point B crosses the dashed hysteresis border reaching the left-half domain (at $t^* = 0.8$), the drying becomes and remains symmetrical until drying completion (SM05). As shown in Fig. \ref{Fig03}d, the contact angle at contact point B changes repetitively between ${\theta _{eq}} = 60^\circ $ and ${\theta _{R,R}} = 30^\circ $ before $t^* = 0.8$, while the contact angle at contact point A only fluctuates within a small range ($ < 6^\circ $). The reason of the latter is attributed to the fast receding of contact point B, providing insufficient time for contact point A to recover to initial contact angle ${\theta _{eq}} = 60^\circ $. After $t^* = 0.8$, the contact angle curves of contact point A and B coincide indicating symmetric drying afterwards. For nanoparticles, Fig. \ref{Fig03}e shows a very high deposition at initial contact point ${A_0}$ in case b, caused by the long duration of capillary flow towards this point before it depins, as shown in Fig. \ref{Fig03}f. Besides, the deposition peak of case b is about 4 times of case a, indicating the unsymmetrical stick-slip processes promote the nanoparticle accumulation, compared to symmetrical ones.

\begin{figure*}[!htbp]
	\center {
		{\epsfig{file=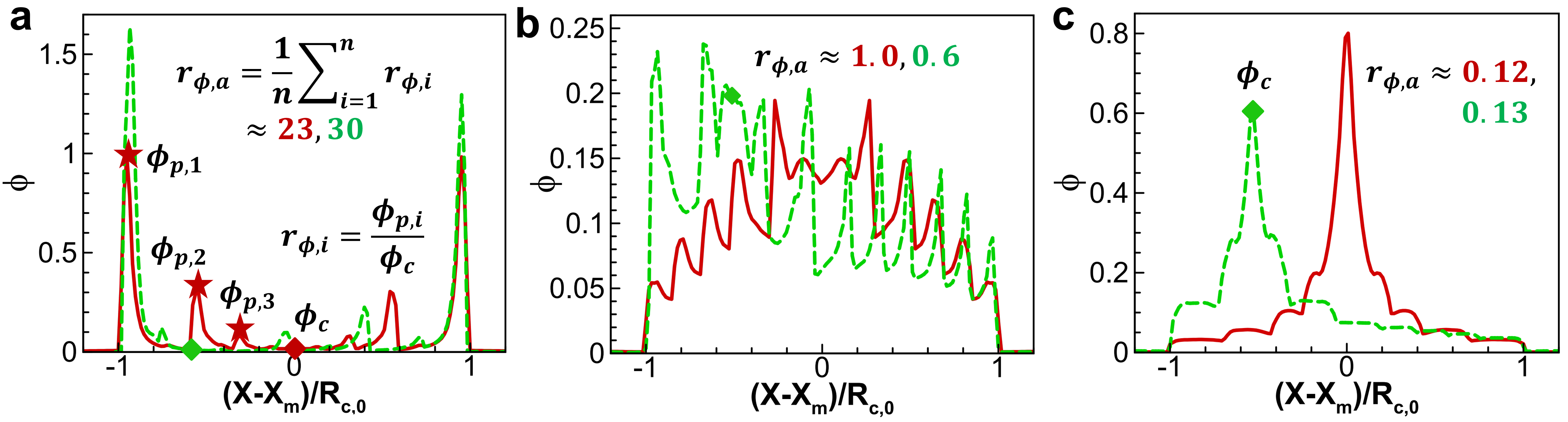,width=0.9\textwidth,clip=}}\hspace{0.0cm} \\ 
	}
	\caption{Different deposition patterns after colloidal droplet drying with consecutive stick-slip processes, where the red-solid and green-dashed curves correspond to uniform and nonuniform contact angle hysteresis ranges given in Figure \ref{Fig03}a and b. (a) Symmetrical and unsymmetrical multiple ring depositions (SM06-07) at ${v_l} = {\rm{ }}1.7 \times {10^{ - 2}}$, $\sigma  = {\rm{ }}7.8 \times {10^{ - 3}}$ and ${D_p} = 1.5 \times {10^{ - 3}}$. (b) Approximately uniform depositions (SM08-09) at ${v_l} = {\rm{ }}1.7 \times {10^{ - 1}}$, $\sigma  = {\rm{ }}3.8 \times {10^{ - 3}}$ and ${D_p} = 1.5 \times {10^{ - 3}}$. (c) Multiple mountain-like depositions (SM10-11) at ${v_l} = {\rm{ }}3.3 \times {10^{ - 1}}$, $\sigma  = {\rm{ }}1.5 \times {10^{ - 2}}$ and ${D_p} = 1.5 \times {10^{ - 2}}$. }
	\label{Fig04}
\end{figure*}

Furthermore, we show that liquid/nanoparticle properties can alter the final deposition pattern obtained after consecutive stick-slip processes. Similar to Fig. \ref{Fig02}, we change the liquid viscosity ${v_l}$, surface tension $\sigma $ and nanoparticle diffusion coefficient ${D_p}$. The contact angle hysteresis conditions are set the same as in Fig. \ref{Fig03}a-b. We use the average deposition ratio ${r_{\phi ,a}} = \frac{1}{n}\sum\nolimits_{i = 1}^n {{r_{\phi ,i}}} $ to quantify various deposition patterns, where $i$ represents the ${i^{th}}$ stick point. Fig. \ref{Fig04} shows 3 different deposition patterns ranging from the multiple ring depositions (Fig. \ref{Fig04}a with ${r_{\phi ,a}} \sim O(10)$) to approximately-uniform (Fig. \ref{Fig04}b with ${r_{\phi ,a}} \sim O(1)$) and mountain-like depositions (Fig. \ref{Fig04}c with ${r_{\phi ,a}} \sim O(0.1)$), regardless of the contact angle hysteresis uniformity. The mechanisms of various depositions in Fig. \ref{Fig04} are similar to those in Fig. \ref{Fig02}.

\begin{figure}[!htbp]
	\center {
		{\epsfig{file=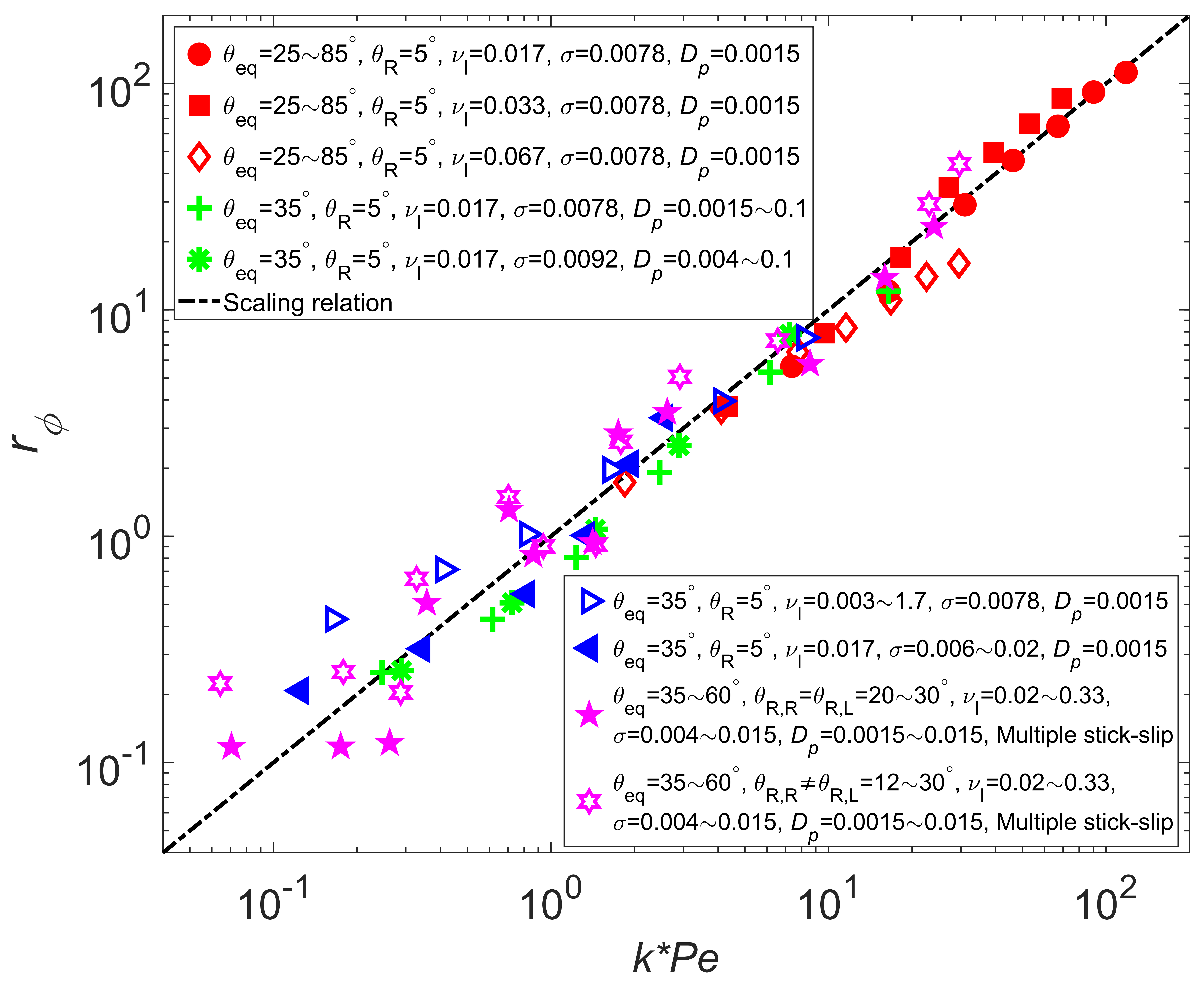,width=0.47\textwidth,clip=}}\hspace{0.0cm}  
	}
	\caption{The linear relation ${r_\phi } \approx k * Pe{\rm{ }},{\rm{ }}k = 6.7 \times {10^{ - 3}}$ between deposition ratio ${r_\phi }$ (nanoparticles at droplet peripheries over nanoparticles at center) and estimated effective Péclet number $Pe$ for a broad range of parameters, i.e., droplet initial contact angle ${\theta _0} \in (25^\circ ,85^\circ )$, droplet receding contact angle ${\theta _R} \in (5^\circ ,30^\circ )$, liquid kinematic viscosity ${v_l} \in (1.7 \times {10^{ - 2}},1.7)$, surface tension $\sigma  \in (6 \times {10^{ - 4}},2 \times {10^{ - 2}})$ and nanoparticle diffusion coefficient ${D_p} \in (1.5 \times {10^{ - 3}},0.1)$.}
	\label{Fig05}
\end{figure}

As we explained in Fig. \ref{Fig02}, the two mechanisms account for the nanoparticle transport during isothermal evaporation are droplet internal capillary flow and nanoparticle diffusion, which determine the deposition ratio. In the following, we investigate and quantify the competition of the two processes considering all patterns studied above, by using the dimensionless Péclet number $Pe = \frac{{{u_c}{R_{c,0}}}}{{{D_p}}}$ representing their strength ratio. ${u_c}$ is the characteristic fluid velocity due to internal capillary transport and ${R_{c,0}}$ is the initial contact radius. The estimation of ${u_c}$ is based on lubrication theory \cite{Kaplan2014,Oron1997}. We define the droplet aspect ratio $\varepsilon  = h/{r_c}$, where $h$ and ${r_c}$ are the droplet height and contact radius during the drying process, respectively. Under a thin film assumption $\varepsilon  \le 0.1$, the balance between radial pressure gradient and viscous force can be written as $\partial p/\partial {r_c} = {\rho _l}{v_l}{\partial ^2}u/\partial {h^2}$. However, when the droplet contact angle is high leading to a high aspect ratio ($\varepsilon  > 0.1$), the viscous force in vertical direction has also to be considered, i.e., $\partial p/\partial {r_c} = {\rho _l}{v_l}({\partial ^2}u/\partial {h^2} + {\partial ^2}u/\partial r_c^2)$. Using Laplace’s law, we obtain the pressure $p \approx \sigma {\rm{ }}{\partial ^2}h/\partial r_c^2$. The two corresponding scaling relations are $p/{r_c} \sim {\rho _l}{v_l}({u_c}/{h^2} + {u_c}/r_c^2)$ and $p \sim \sigma h/r_c^2$, and their combination makes the scaled characteristic velocity ${u_c} \sim \frac{\sigma }{{{\rho _l}{v_l}}}\frac{{{\varepsilon ^3}}}{{1 + {\varepsilon ^2}}}$. Using as a simplification the initial aspect ratio ${\varepsilon _0} = {H_0}/{R_{c,0}}$, we obtain the effective Péclet number as
\begin{equation}\label{eq:Pe}
	Pe = \frac{{{u_c}{R_{c,0}}}}{{{D_p}}} \sim \frac{{\sigma {R_{c,0}}}}{{{\rho _l}{v_l}{D_p}}}\frac{{\varepsilon _0^3}}{{1 + \varepsilon _0^2}}.
\end{equation}
Eq.(\ref{eq:Pe}) shows that high liquid-vapor surface tension $\sigma $, high initial aspect ratio ${\varepsilon _0}$, low liquid viscosity ${v_l}$, low nanoparticle diffusion coefficient ${D_p}$ lead to high $Pe$. We performed an extensive parametric analysis spanning ${v_l} \in (1.7 \times {10^{ - 2}},1.7)$, $\sigma  \in (6 \times {10^{ - 4}},2 \times {10^{ - 2}})$, ${D_p} \in (1.5 \times {10^{ - 3}},0.1)$, and ${\varepsilon _0} \in (0.24,0.93)$ [corresponding to ${\theta _0} \in (25^\circ ,85^\circ )$]. The deposition curves for the different parameters are shown in Fig. S7. The coffee ring is preferably formed at high $\sigma $, low ${v_l}$ and small ${D_p}$, while mountain-like depositions are observed at opposite parameter limits. Uniform deposition occurs at intermediate conditions. To investigate the relation between $Pe$ and the final nanoparticle deposition pattern quantitatively, we plot the deposition ratio ${r_\phi }$ against $Pe$. Fig. \ref{Fig05} shows a very neat linear relation (the slope is 1 for the log-log axes):
\begin{equation}\label{eq:scaling}
	{r_\phi } \approx k * Pe,{\rm{ }}k = 6.7 \times {10^{ - 3}}
\end{equation}
within a large range of $Pe$ values covering over 3 orders of magnitude. The prefactor $k \ll 1$ indicates that very intense capillary flow is required to transport the nanoparticles at the pinned contact point. The scaling relation is further extended to depositions after consecutive stick-slip processes as shown in Fig. \ref{Fig03} and Fig. \ref{Fig04}. For each stick-slip process ($i^{th}$ for instance), we first calculate two Péclet numbers at the beginning of stick phase ($P{e_{i,sk}}$) and slip phase ($P{e_{i,sp}}$) with Eq.(\ref{eq:Pe}). Then we use their subtraction ($P{e_i} = P{e_{i,sk}} - P{e_{i,sp}}$) to quantify the nanoparticle motion in the $i^th$ stick-slip phase. To consider all stick-slip processes during drying, we simply use the average value of $P{e_i} $ as the effective Peclect number, i.e., $P{e_a} = \frac{1}{n}\sum\nolimits_{i = 1}^n {P{e_i}} $. More explanations are given in Fig. S8 in Supplementary Files. Note that unsymmetrical drying and deposition processes can be similarly treated. Extensive simulations spanning ${v_l} \in (1.7 \times {10^{ - 2}},3.3 \times {10^{ - 1}})$, $\sigma  \in (3.8 \times {10^{ - 3}},1.5 \times {10^{ - 2}})$, ${D_p} \in (1.5 \times {10^{ - 3}},1.5 \times {10^{ - 2}})$, ${\theta _0} \in (35^\circ ,60^\circ )$ and ${\theta _{R,L}},{\theta _{R,R}} \in (12^\circ ,30^\circ )$ are conducted, with the modeled deposition profiles shown in Fig. S9-10. The comparison between the average deposition ratio ${r_{\phi ,a}}$ and average effective Péclet number $P{e_a}$ after symmetrical/ unsymmetrical multiple stick-slip processes is also given in Fig. \ref{Fig05}. Note that, for expression simplicity and uniformity, we still use ${r_\phi }$ and $Pe$ instead of ${r_{\phi ,a}}$ and $P{e_a}$ in the figure. Generally, a good agreement is achieved. Some deviations are observed, mainly because droplet profiles are not exactly circular when experiencing consecutive stick-slip, affecting the estimation of pressure and further the $Pe$. For all the studied cases from simple mountain-like, uniform, single ring to un-/symmetrical multiple rings, the overall coefficient of determination is up to ${R^2} = 0.96$, indicating very good reliability to estimate ${r_\phi }$ by Eq.(\ref{eq:scaling}) through the effective Péclet number in Eq.(\ref{eq:Pe}).

To conclude, based on extensive numerical simulations and theoretical analysis, we proposed a scaling relation for the colloidal droplet evaporation which can quantitatively describe the final nanoparticle deposition ratio for various patterns, from ring, uniform to mountain-like depositions, experiencing single or multiple symmetrical/unsymmetrical stick-slip processes. It confirms that complex nanoparticle deposition behaviors are elegantly controlled by the balance of capillary convection and diffusion of nanoparticles through an effective Péclet number. The effective Péclet number is a priori and concisely defined on the solvent/nanoparticle/substrate properties, which therefore can be potentially applied in real applications to guide nanoparticle depositions like ink-jet printing and nanomaterial fabrication, to name but a few. What remains to be studied, among others, is the Marangoni effects on nanoparticle transport caused by temperature variation or surfactant gradient.

\section{Acknowledgements}
This work is supported by Swiss National Science Foundation (SNF, project No. 175793) and the Fundamental Research Funds for the Central Universities (G2022KY05105). Computational support is provided by Swiss National Super Computing Center (Project No. s1081). Q. K. acknowledges LANL’s LDRD Program. S. S. is funded by the European Research Council under the European Unions Horizon 2020 Framework Programme (No. FP/2014-2020)/ERC Grant Agreement No. 739964 (COPMAT).
\nocite{*}
\bibliography{GMRT}

\begin{thebibliography}{36}%
\makeatletter
\providecommand \@ifxundefined [1]{%
 \@ifx{#1\undefined}
}%
\providecommand \@ifnum [1]{%
 \ifnum #1\expandafter \@firstoftwo
 \else \expandafter \@secondoftwo
 \fi
}%
\providecommand \@ifx [1]{%
 \ifx #1\expandafter \@firstoftwo
 \else \expandafter \@secondoftwo
 \fi
}%
\providecommand \natexlab [1]{#1}%
\providecommand \enquote  [1]{``#1''}%
\providecommand \bibnamefont  [1]{#1}%
\providecommand \bibfnamefont [1]{#1}%
\providecommand \citenamefont [1]{#1}%
\providecommand \href@noop [0]{\@secondoftwo}%
\providecommand \href [0]{\begingroup \@sanitize@url \@href}%
\providecommand \@href[1]{\@@startlink{#1}\@@href}%
\providecommand \@@href[1]{\endgroup#1\@@endlink}%
\providecommand \@sanitize@url [0]{\catcode `\\12\catcode `\$12\catcode
  `\&12\catcode `\#12\catcode `\^12\catcode `\_12\catcode `\%12\relax}%
\providecommand \@@startlink[1]{}%
\providecommand \@@endlink[0]{}%
\providecommand \url  [0]{\begingroup\@sanitize@url \@url }%
\providecommand \@url [1]{\endgroup\@href {#1}{\urlprefix }}%
\providecommand \urlprefix  [0]{URL }%
\providecommand \Eprint [0]{\href }%
\providecommand \doibase [0]{https://doi.org/}%
\providecommand \selectlanguage [0]{\@gobble}%
\providecommand \bibinfo  [0]{\@secondoftwo}%
\providecommand \bibfield  [0]{\@secondoftwo}%
\providecommand \translation [1]{[#1]}%
\providecommand \BibitemOpen [0]{}%
\providecommand \bibitemStop [0]{}%
\providecommand \bibitemNoStop [0]{.\EOS\space}%
\providecommand \EOS [0]{\spacefactor3000\relax}%
\providecommand \BibitemShut  [1]{\csname bibitem#1\endcsname}%
\let\auto@bib@innerbib\@empty
\bibitem [{\citenamefont {Bonn}\ \emph {et~al.}(2009)\citenamefont {Bonn},
  \citenamefont {Eggers}, \citenamefont {Indekeu},\ and\ \citenamefont
  {Meunier}}]{Bonn2009}%
  \BibitemOpen
  \bibfield  {author} {\bibinfo {author} {\bibfnamefont {D.}~\bibnamefont
  {Bonn}}, \bibinfo {author} {\bibfnamefont {J.}~\bibnamefont {Eggers}},
  \bibinfo {author} {\bibfnamefont {J.}~\bibnamefont {Indekeu}},\ and\ \bibinfo
  {author} {\bibfnamefont {J.}~\bibnamefont {Meunier}},\ }\bibfield  {title}
  {\bibinfo {title} {{Wetting and spreading}},\ }\href
  {https://doi.org/10.1103/RevModPhys.81.739} {\bibfield  {journal} {\bibinfo
  {journal} {Reviews of Modern Physics}\ }\textbf {\bibinfo {volume} {81}},\
  \bibinfo {pages} {739} (\bibinfo {year} {2009})}\BibitemShut {NoStop}%
\bibitem [{\citenamefont {Mampallil}\ and\ \citenamefont
  {Eral}(2018)}]{Mampallil2018}%
  \BibitemOpen
  \bibfield  {author} {\bibinfo {author} {\bibfnamefont {D.}~\bibnamefont
  {Mampallil}}\ and\ \bibinfo {author} {\bibfnamefont {H.~B.}\ \bibnamefont
  {Eral}},\ }\bibfield  {title} {\bibinfo {title} {{A review on suppression and
  utilization of the coffee-ring effect}},\ }\href
  {https://doi.org/10.1016/j.cis.2017.12.008} {\bibfield  {journal} {\bibinfo
  {journal} {Advances in Colloid and Interface Science}\ }\textbf {\bibinfo
  {volume} {252}},\ \bibinfo {pages} {38} (\bibinfo {year} {2018})}\BibitemShut
  {NoStop}%
\bibitem [{\citenamefont {Deegan}\ \emph {et~al.}(1997)\citenamefont {Deegan},
  \citenamefont {Bakajin}, \citenamefont {Dupont}, \citenamefont {Huber},
  \citenamefont {Nagel},\ and\ \citenamefont {Witten}}]{Deegan1997}%
  \BibitemOpen
  \bibfield  {author} {\bibinfo {author} {\bibfnamefont {R.~D.}\ \bibnamefont
  {Deegan}}, \bibinfo {author} {\bibfnamefont {O.}~\bibnamefont {Bakajin}},
  \bibinfo {author} {\bibfnamefont {T.~F.}\ \bibnamefont {Dupont}}, \bibinfo
  {author} {\bibfnamefont {G.}~\bibnamefont {Huber}}, \bibinfo {author}
  {\bibfnamefont {S.~R.}\ \bibnamefont {Nagel}},\ and\ \bibinfo {author}
  {\bibfnamefont {T.~A.}\ \bibnamefont {Witten}},\ }\bibfield  {title}
  {\bibinfo {title} {{Capillary flow as the cause of ring stains from dried
  liquid drops}},\ }\href {https://doi.org/10.1038/39827} {\bibfield  {journal}
  {\bibinfo  {journal} {Nature}\ }\textbf {\bibinfo {volume} {389}},\ \bibinfo
  {pages} {827} (\bibinfo {year} {1997})}\BibitemShut {NoStop}%
\bibitem [{\citenamefont {Shmuylovich}\ \emph {et~al.}(2002)\citenamefont
  {Shmuylovich}, \citenamefont {Shen},\ and\ \citenamefont
  {Stone}}]{Shmuylovich2002}%
  \BibitemOpen
  \bibfield  {author} {\bibinfo {author} {\bibfnamefont {L.}~\bibnamefont
  {Shmuylovich}}, \bibinfo {author} {\bibfnamefont {A.~Q.}\ \bibnamefont
  {Shen}},\ and\ \bibinfo {author} {\bibfnamefont {H.~A.}\ \bibnamefont
  {Stone}},\ }\bibfield  {title} {\bibinfo {title} {{Surface morphology of
  drying latex films: Multiple ring formation}},\ }\href
  {https://doi.org/10.1021/la011484v} {\bibfield  {journal} {\bibinfo
  {journal} {Langmuir}\ }\textbf {\bibinfo {volume} {18}},\ \bibinfo {pages}
  {3441} (\bibinfo {year} {2002})}\BibitemShut {NoStop}%
\bibitem [{\citenamefont {Maheshwari}\ \emph {et~al.}(2008)\citenamefont
  {Maheshwari}, \citenamefont {Zhang}, \citenamefont {Zhu},\ and\ \citenamefont
  {Chang}}]{Maheshwari2008}%
  \BibitemOpen
  \bibfield  {author} {\bibinfo {author} {\bibfnamefont {S.}~\bibnamefont
  {Maheshwari}}, \bibinfo {author} {\bibfnamefont {L.}~\bibnamefont {Zhang}},
  \bibinfo {author} {\bibfnamefont {Y.}~\bibnamefont {Zhu}},\ and\ \bibinfo
  {author} {\bibfnamefont {H.~C.}\ \bibnamefont {Chang}},\ }\bibfield  {title}
  {\bibinfo {title} {{Coupling between precipitation and contact-line dynamics:
  Multiring stains and stick-slip motion}},\ }\href
  {https://doi.org/10.1103/PhysRevLett.100.044503} {\bibfield  {journal}
  {\bibinfo  {journal} {Physical Review Letters}\ }\textbf {\bibinfo {volume}
  {100}},\ \bibinfo {pages} {1} (\bibinfo {year} {2008})}\BibitemShut {NoStop}%
\bibitem [{\citenamefont {Moffat}\ \emph {et~al.}(2009)\citenamefont {Moffat},
  \citenamefont {Sefiane},\ and\ \citenamefont {Shanahan}}]{Moffat2009}%
  \BibitemOpen
  \bibfield  {author} {\bibinfo {author} {\bibfnamefont {J.~R.}\ \bibnamefont
  {Moffat}}, \bibinfo {author} {\bibfnamefont {K.}~\bibnamefont {Sefiane}},\
  and\ \bibinfo {author} {\bibfnamefont {M.~E.}\ \bibnamefont {Shanahan}},\
  }\bibfield  {title} {\bibinfo {title} {{Effect of TiO2 nanoparticles on
  contact line stick-slip behavior of volatile drops}},\ }\href
  {https://doi.org/10.1021/jp902062z} {\bibfield  {journal} {\bibinfo
  {journal} {Journal of Physical Chemistry B}\ }\textbf {\bibinfo {volume}
  {113}},\ \bibinfo {pages} {8860} (\bibinfo {year} {2009})}\BibitemShut
  {NoStop}%
\bibitem [{\citenamefont {Yang}\ \emph {et~al.}(2014)\citenamefont {Yang},
  \citenamefont {Li},\ and\ \citenamefont {Sun}}]{Yang2014}%
  \BibitemOpen
  \bibfield  {author} {\bibinfo {author} {\bibfnamefont {X.}~\bibnamefont
  {Yang}}, \bibinfo {author} {\bibfnamefont {C.~Y.}\ \bibnamefont {Li}},\ and\
  \bibinfo {author} {\bibfnamefont {Y.}~\bibnamefont {Sun}},\ }\bibfield
  {title} {\bibinfo {title} {{From multi-ring to spider web and radial spoke:
  Competition between the receding contact line and particle deposition in a
  drying colloidal drop}},\ }\href {https://doi.org/10.1039/c4sm00497c}
  {\bibfield  {journal} {\bibinfo  {journal} {Soft Matter}\ }\textbf {\bibinfo
  {volume} {10}},\ \bibinfo {pages} {4458} (\bibinfo {year}
  {2014})}\BibitemShut {NoStop}%
\bibitem [{\citenamefont {Willmer}\ \emph {et~al.}(2010)\citenamefont
  {Willmer}, \citenamefont {Baldwin}, \citenamefont {Kwartnik},\ and\
  \citenamefont {Fairhurst}}]{Willmer2010}%
  \BibitemOpen
  \bibfield  {author} {\bibinfo {author} {\bibfnamefont {D.}~\bibnamefont
  {Willmer}}, \bibinfo {author} {\bibfnamefont {K.~A.}\ \bibnamefont
  {Baldwin}}, \bibinfo {author} {\bibfnamefont {C.}~\bibnamefont {Kwartnik}},\
  and\ \bibinfo {author} {\bibfnamefont {D.~J.}\ \bibnamefont {Fairhurst}},\
  }\bibfield  {title} {\bibinfo {title} {{Growth of solid conical structures
  during multistage drying of sessile poly(ethylene oxide) droplets}},\ }\href
  {https://doi.org/10.1039/b922727j} {\bibfield  {journal} {\bibinfo  {journal}
  {Physical Chemistry Chemical Physics}\ }\textbf {\bibinfo {volume} {12}},\
  \bibinfo {pages} {3998} (\bibinfo {year} {2010})}\BibitemShut {NoStop}%
\bibitem [{\citenamefont {Li}\ \emph {et~al.}(2014)\citenamefont {Li},
  \citenamefont {Sheng},\ and\ \citenamefont {Tsao}}]{Li2014c}%
  \BibitemOpen
  \bibfield  {author} {\bibinfo {author} {\bibfnamefont {Y.~F.}\ \bibnamefont
  {Li}}, \bibinfo {author} {\bibfnamefont {Y.~J.}\ \bibnamefont {Sheng}},\ and\
  \bibinfo {author} {\bibfnamefont {H.~K.}\ \bibnamefont {Tsao}},\ }\bibfield
  {title} {\bibinfo {title} {{Solute concentration-dependent contact angle
  hysteresis and evaporation stains}},\ }\href
  {https://doi.org/10.1021/la501438k} {\bibfield  {journal} {\bibinfo
  {journal} {Langmuir}\ }\textbf {\bibinfo {volume} {30}},\ \bibinfo {pages}
  {7716} (\bibinfo {year} {2014})}\BibitemShut {NoStop}%
\bibitem [{\citenamefont {Yunker}\ \emph {et~al.}(2011)\citenamefont {Yunker},
  \citenamefont {Still}, \citenamefont {Lohr},\ and\ \citenamefont
  {Yodh}}]{Yunker2011}%
  \BibitemOpen
  \bibfield  {author} {\bibinfo {author} {\bibfnamefont {P.~J.}\ \bibnamefont
  {Yunker}}, \bibinfo {author} {\bibfnamefont {T.}~\bibnamefont {Still}},
  \bibinfo {author} {\bibfnamefont {M.~A.}\ \bibnamefont {Lohr}},\ and\
  \bibinfo {author} {\bibfnamefont {A.~G.}\ \bibnamefont {Yodh}},\ }\bibfield
  {title} {\bibinfo {title} {{Suppression of the coffee-ring effect by
  shape-dependent capillary interactions}},\ }\href
  {https://doi.org/10.1038/nature10344} {\bibfield  {journal} {\bibinfo
  {journal} {Nature}\ }\textbf {\bibinfo {volume} {476}},\ \bibinfo {pages}
  {308} (\bibinfo {year} {2011})}\BibitemShut {NoStop}%
\bibitem [{\citenamefont {Bhardwaj}\ \emph {et~al.}(2010)\citenamefont
  {Bhardwaj}, \citenamefont {Fang}, \citenamefont {Somasundaran},\ and\
  \citenamefont {Attinger}}]{Bhardwaj2010}%
  \BibitemOpen
  \bibfield  {author} {\bibinfo {author} {\bibfnamefont {R.}~\bibnamefont
  {Bhardwaj}}, \bibinfo {author} {\bibfnamefont {X.}~\bibnamefont {Fang}},
  \bibinfo {author} {\bibfnamefont {P.}~\bibnamefont {Somasundaran}},\ and\
  \bibinfo {author} {\bibfnamefont {D.}~\bibnamefont {Attinger}},\ }\bibfield
  {title} {\bibinfo {title} {{Self-assembly of colloidal particles from
  evaporating droplets: Role of DLVO interactions and proposition of a phase
  diagram}},\ }\href {https://doi.org/10.1021/la9047227} {\bibfield  {journal}
  {\bibinfo  {journal} {Langmuir}\ }\textbf {\bibinfo {volume} {26}},\ \bibinfo
  {pages} {7833} (\bibinfo {year} {2010})},\ \Eprint
  {https://arxiv.org/abs/1010.2564} {arXiv:1010.2564} \BibitemShut {NoStop}%
\bibitem [{\citenamefont {Kajiya}\ \emph {et~al.}(2009)\citenamefont {Kajiya},
  \citenamefont {Monteux}, \citenamefont {Narita}, \citenamefont {Lequeux},\
  and\ \citenamefont {Doi}}]{Kajiya2009}%
  \BibitemOpen
  \bibfield  {author} {\bibinfo {author} {\bibfnamefont {T.}~\bibnamefont
  {Kajiya}}, \bibinfo {author} {\bibfnamefont {C.}~\bibnamefont {Monteux}},
  \bibinfo {author} {\bibfnamefont {T.}~\bibnamefont {Narita}}, \bibinfo
  {author} {\bibfnamefont {F.}~\bibnamefont {Lequeux}},\ and\ \bibinfo {author}
  {\bibfnamefont {M.}~\bibnamefont {Doi}},\ }\bibfield  {title} {\bibinfo
  {title} {{Contact-line recession leaving a macroscopic polymer film in the
  drying droplets of water-poly(N, N-dimethylacrylamide) (PDMA) solution}},\
  }\href {https://doi.org/10.1021/la900216k} {\bibfield  {journal} {\bibinfo
  {journal} {Langmuir}\ }\textbf {\bibinfo {volume} {25}},\ \bibinfo {pages}
  {6934} (\bibinfo {year} {2009})}\BibitemShut {NoStop}%
\bibitem [{\citenamefont {Truskett}\ and\ \citenamefont
  {Stebe}(2003)}]{Truskett2003}%
  \BibitemOpen
  \bibfield  {author} {\bibinfo {author} {\bibfnamefont {V.~N.}\ \bibnamefont
  {Truskett}}\ and\ \bibinfo {author} {\bibfnamefont {K.~J.}\ \bibnamefont
  {Stebe}},\ }\bibfield  {title} {\bibinfo {title} {{Influence of surfactants
  on an evaporating drop: Fluorescence images and particle deposition
  patterns}},\ }\href {https://doi.org/10.1021/la030049t} {\bibfield  {journal}
  {\bibinfo  {journal} {Langmuir}\ }\textbf {\bibinfo {volume} {19}},\ \bibinfo
  {pages} {8271} (\bibinfo {year} {2003})}\BibitemShut {NoStop}%
\bibitem [{\citenamefont {Deegan}\ \emph {et~al.}(2000)\citenamefont {Deegan},
  \citenamefont {Bakajin}, \citenamefont {Dupont}, \citenamefont {Huber},
  \citenamefont {Nagel},\ and\ \citenamefont {Witten}}]{Deegan2000a}%
  \BibitemOpen
  \bibfield  {author} {\bibinfo {author} {\bibfnamefont {R.~D.}\ \bibnamefont
  {Deegan}}, \bibinfo {author} {\bibfnamefont {O.}~\bibnamefont {Bakajin}},
  \bibinfo {author} {\bibfnamefont {T.~F.}\ \bibnamefont {Dupont}}, \bibinfo
  {author} {\bibfnamefont {G.}~\bibnamefont {Huber}}, \bibinfo {author}
  {\bibfnamefont {S.~R.}\ \bibnamefont {Nagel}},\ and\ \bibinfo {author}
  {\bibfnamefont {T.~A.}\ \bibnamefont {Witten}},\ }\bibfield  {title}
  {\bibinfo {title} {{Contact line deposits in an evaporating drop}},\ }\href
  {https://doi.org/10.1103/PhysRevE.62.756} {\bibfield  {journal} {\bibinfo
  {journal} {Physical Review E - Statistical Physics, Plasmas, Fluids, and
  Related Interdisciplinary Topics}\ }\textbf {\bibinfo {volume} {62}},\
  \bibinfo {pages} {756} (\bibinfo {year} {2000})}\BibitemShut {NoStop}%
\bibitem [{\citenamefont {Hu}\ and\ \citenamefont {Larson}(2002)}]{Hu2002}%
  \BibitemOpen
  \bibfield  {author} {\bibinfo {author} {\bibfnamefont {H.}~\bibnamefont
  {Hu}}\ and\ \bibinfo {author} {\bibfnamefont {R.~G.}\ \bibnamefont
  {Larson}},\ }\bibfield  {title} {\bibinfo {title} {{Evaporation of a sessile
  droplet on a substrate}},\ }\href {https://doi.org/10.1021/jp0118322}
  {\bibfield  {journal} {\bibinfo  {journal} {Journal of Physical Chemistry B}\
  }\textbf {\bibinfo {volume} {106}},\ \bibinfo {pages} {1334} (\bibinfo {year}
  {2002})}\BibitemShut {NoStop}%
\bibitem [{\citenamefont {Hu}\ and\ \citenamefont
  {Larson}(2005{\natexlab{a}})}]{Hu2005b}%
  \BibitemOpen
  \bibfield  {author} {\bibinfo {author} {\bibfnamefont {H.}~\bibnamefont
  {Hu}}\ and\ \bibinfo {author} {\bibfnamefont {R.~G.}\ \bibnamefont
  {Larson}},\ }\bibfield  {title} {\bibinfo {title} {{Analysis of the
  microfluid flow in an evaporating sessile droplet}},\ }\href
  {https://doi.org/10.1021/la047528s} {\bibfield  {journal} {\bibinfo
  {journal} {Langmuir}\ }\textbf {\bibinfo {volume} {21}},\ \bibinfo {pages}
  {3963} (\bibinfo {year} {2005}{\natexlab{a}})}\BibitemShut {NoStop}%
\bibitem [{\citenamefont {Hu}\ and\ \citenamefont
  {Larson}(2005{\natexlab{b}})}]{Hu2005a}%
  \BibitemOpen
  \bibfield  {author} {\bibinfo {author} {\bibfnamefont {H.}~\bibnamefont
  {Hu}}\ and\ \bibinfo {author} {\bibfnamefont {R.~G.}\ \bibnamefont
  {Larson}},\ }\bibfield  {title} {\bibinfo {title} {{Analysis of the effects
  of marangoni stresses on the microflow in an evaporating sessile droplet}},\
  }\href {https://doi.org/10.1021/la0475270} {\bibfield  {journal} {\bibinfo
  {journal} {Langmuir}\ }\textbf {\bibinfo {volume} {21}},\ \bibinfo {pages}
  {3972} (\bibinfo {year} {2005}{\natexlab{b}})}\BibitemShut {NoStop}%
\bibitem [{\citenamefont {Bhardwaj}\ \emph {et~al.}(2009)\citenamefont
  {Bhardwaj}, \citenamefont {Fang},\ and\ \citenamefont
  {Attinger}}]{Bhardwaj2009a}%
  \BibitemOpen
  \bibfield  {author} {\bibinfo {author} {\bibfnamefont {R.}~\bibnamefont
  {Bhardwaj}}, \bibinfo {author} {\bibfnamefont {X.}~\bibnamefont {Fang}},\
  and\ \bibinfo {author} {\bibfnamefont {D.}~\bibnamefont {Attinger}},\
  }\bibfield  {title} {\bibinfo {title} {{Pattern formation during the
  evaporation of a colloidal nanoliter drop: A numerical and experimental
  study}},\ }\bibfield  {journal} {\bibinfo  {journal} {New Journal of
  Physics}\ }\textbf {\bibinfo {volume} {11}},\ \href
  {https://doi.org/10.1088/1367-2630/11/7/075020}
  {10.1088/1367-2630/11/7/075020} (\bibinfo {year} {2009})\BibitemShut
  {NoStop}%
\bibitem [{\citenamefont {Frastia}\ \emph {et~al.}(2011)\citenamefont
  {Frastia}, \citenamefont {Archer},\ and\ \citenamefont
  {Thiele}}]{Frastia2011}%
  \BibitemOpen
  \bibfield  {author} {\bibinfo {author} {\bibfnamefont {L.}~\bibnamefont
  {Frastia}}, \bibinfo {author} {\bibfnamefont {A.~J.}\ \bibnamefont
  {Archer}},\ and\ \bibinfo {author} {\bibfnamefont {U.}~\bibnamefont
  {Thiele}},\ }\bibfield  {title} {\bibinfo {title} {{Dynamical model for the
  formation of patterned deposits at receding contact lines}},\ }\href
  {https://doi.org/10.1103/PhysRevLett.106.077801} {\bibfield  {journal}
  {\bibinfo  {journal} {Physical Review Letters}\ }\textbf {\bibinfo {volume}
  {106}},\ \bibinfo {pages} {1} (\bibinfo {year} {2011})}\BibitemShut {NoStop}%
\bibitem [{\citenamefont {Man}\ and\ \citenamefont {Doi}(2016)}]{Man2016}%
  \BibitemOpen
  \bibfield  {author} {\bibinfo {author} {\bibfnamefont {X.}~\bibnamefont
  {Man}}\ and\ \bibinfo {author} {\bibfnamefont {M.}~\bibnamefont {Doi}},\
  }\bibfield  {title} {\bibinfo {title} {{Ring to Mountain Transition in
  Deposition Pattern of Drying Droplets}},\ }\href
  {https://doi.org/10.1103/PhysRevLett.116.066101} {\bibfield  {journal}
  {\bibinfo  {journal} {Physical Review Letters}\ }\textbf {\bibinfo {volume}
  {116}},\ \bibinfo {pages} {066101} (\bibinfo {year} {2016})}\BibitemShut
  {NoStop}%
\bibitem [{\citenamefont {Wu}\ \emph {et~al.}(2018)\citenamefont {Wu},
  \citenamefont {Man},\ and\ \citenamefont {Doi}}]{Wu2018}%
  \BibitemOpen
  \bibfield  {author} {\bibinfo {author} {\bibfnamefont {M.}~\bibnamefont
  {Wu}}, \bibinfo {author} {\bibfnamefont {X.}~\bibnamefont {Man}},\ and\
  \bibinfo {author} {\bibfnamefont {M.}~\bibnamefont {Doi}},\ }\bibfield
  {title} {\bibinfo {title} {{Multi-ring Deposition Pattern of Drying
  Droplets}},\ }\href {https://doi.org/10.1021/acs.langmuir.8b01655} {\bibfield
   {journal} {\bibinfo  {journal} {Langmuir}\ }\textbf {\bibinfo {volume}
  {34}},\ \bibinfo {pages} {9572} (\bibinfo {year} {2018})},\ \Eprint
  {https://arxiv.org/abs/1807.09673} {arXiv:1807.09673} \BibitemShut {NoStop}%
\bibitem [{\citenamefont {Kaplan}\ and\ \citenamefont
  {Mahadevan}(2015)}]{Kaplan2014}%
  \BibitemOpen
  \bibfield  {author} {\bibinfo {author} {\bibfnamefont {C.~N.}\ \bibnamefont
  {Kaplan}}\ and\ \bibinfo {author} {\bibfnamefont {L.}~\bibnamefont
  {Mahadevan}},\ }\bibfield  {title} {\bibinfo {title} {{Evaporation-driven
  ring and film deposition from colloidal droplets}},\ }\href
  {https://doi.org/10.1017/jfm.2015.496} {\bibfield  {journal} {\bibinfo
  {journal} {Journal of Fluid Mechanics}\ }\textbf {\bibinfo {volume} {781}},\
  \bibinfo {pages} {R2} (\bibinfo {year} {2015})},\ \Eprint
  {https://arxiv.org/abs/1411.4748} {arXiv:1411.4748} \BibitemShut {NoStop}%
\bibitem [{\citenamefont {Zhang}\ \emph {et~al.}(2021)\citenamefont {Zhang},
  \citenamefont {Zhang}, \citenamefont {Zhao},\ and\ \citenamefont
  {Yang}}]{Zhang2021a}%
  \BibitemOpen
  \bibfield  {author} {\bibinfo {author} {\bibfnamefont {C.}~\bibnamefont
  {Zhang}}, \bibinfo {author} {\bibfnamefont {H.}~\bibnamefont {Zhang}},
  \bibinfo {author} {\bibfnamefont {Y.}~\bibnamefont {Zhao}},\ and\ \bibinfo
  {author} {\bibfnamefont {C.}~\bibnamefont {Yang}},\ }\bibfield  {title}
  {\bibinfo {title} {{An immersed boundary-lattice Boltzmann model for
  simulation of deposited particle patterns in an evaporating sessile droplet
  with dispersed particles}},\ }\href
  {https://doi.org/10.1016/j.ijheatmasstransfer.2021.121905} {\bibfield
  {journal} {\bibinfo  {journal} {International Journal of Heat and Mass
  Transfer}\ }\textbf {\bibinfo {volume} {181}},\ \bibinfo {pages} {121905}
  (\bibinfo {year} {2021})}\BibitemShut {NoStop}%
\bibitem [{\citenamefont {Zhao}\ and\ \citenamefont {Yong}(2018)}]{Zhao2018b}%
  \BibitemOpen
  \bibfield  {author} {\bibinfo {author} {\bibfnamefont {M.}~\bibnamefont
  {Zhao}}\ and\ \bibinfo {author} {\bibfnamefont {X.}~\bibnamefont {Yong}},\
  }\bibfield  {title} {\bibinfo {title} {{Nanoparticle motion on the surface of
  drying droplets}},\ }\href {https://doi.org/10.1103/PhysRevFluids.3.034201}
  {\bibfield  {journal} {\bibinfo  {journal} {Physical Review Fluids}\ }\textbf
  {\bibinfo {volume} {3}},\ \bibinfo {pages} {1} (\bibinfo {year}
  {2018})}\BibitemShut {NoStop}%
\bibitem [{\citenamefont {Qin}\ \emph {et~al.}(2019{\natexlab{a}})\citenamefont
  {Qin}, \citenamefont {{Mazloomi Moqaddam}}, \citenamefont {{Del Carro}},
  \citenamefont {Kang}, \citenamefont {Brunschwiler}, \citenamefont {Derome},\
  and\ \citenamefont {Carmeliet}}]{Qin2019a}%
  \BibitemOpen
  \bibfield  {author} {\bibinfo {author} {\bibfnamefont {F.}~\bibnamefont
  {Qin}}, \bibinfo {author} {\bibfnamefont {A.}~\bibnamefont {{Mazloomi
  Moqaddam}}}, \bibinfo {author} {\bibfnamefont {L.}~\bibnamefont {{Del
  Carro}}}, \bibinfo {author} {\bibfnamefont {Q.}~\bibnamefont {Kang}},
  \bibinfo {author} {\bibfnamefont {T.}~\bibnamefont {Brunschwiler}}, \bibinfo
  {author} {\bibfnamefont {D.}~\bibnamefont {Derome}},\ and\ \bibinfo {author}
  {\bibfnamefont {J.}~\bibnamefont {Carmeliet}},\ }\bibfield  {title} {\bibinfo
  {title} {{Tricoupled hybrid lattice Boltzmann model for nonisothermal drying
  of colloidal suspensions in micropore structures}},\ }\bibfield  {journal}
  {\bibinfo  {journal} {Physical Review E}\ }\textbf {\bibinfo {volume} {99}},\
  \href {https://doi.org/10.1103/PhysRevE.99.053306}
  {10.1103/PhysRevE.99.053306} (\bibinfo {year}
  {2019}{\natexlab{a}})\BibitemShut {NoStop}%
\bibitem [{\citenamefont {Nath}\ and\ \citenamefont {Ray}(2021)}]{Nath2021}%
  \BibitemOpen
  \bibfield  {author} {\bibinfo {author} {\bibfnamefont {G.}~\bibnamefont
  {Nath}}\ and\ \bibinfo {author} {\bibfnamefont {B.}~\bibnamefont {Ray}},\
  }\bibfield  {title} {\bibinfo {title} {{Manipulating the three-phase contact
  line of an evaporating particle-laden droplet to get desirable
  microstructures: A lattice Boltzmann study}},\ }\href
  {https://doi.org/10.1063/5.0052878} {\bibfield  {journal} {\bibinfo
  {journal} {Physics of Fluids}\ }\textbf {\bibinfo {volume} {33}},\ \bibinfo
  {pages} {083304} (\bibinfo {year} {2021})}\BibitemShut {NoStop}%
\bibitem [{\citenamefont {Li}\ \emph {et~al.}(2016)\citenamefont {Li},
  \citenamefont {Luo}, \citenamefont {Kang}, \citenamefont {He}, \citenamefont
  {Chen},\ and\ \citenamefont {Liu}}]{Li2016d}%
  \BibitemOpen
  \bibfield  {author} {\bibinfo {author} {\bibfnamefont {Q.}~\bibnamefont
  {Li}}, \bibinfo {author} {\bibfnamefont {K.~H.}\ \bibnamefont {Luo}},
  \bibinfo {author} {\bibfnamefont {Q.~J.}\ \bibnamefont {Kang}}, \bibinfo
  {author} {\bibfnamefont {Y.~L.}\ \bibnamefont {He}}, \bibinfo {author}
  {\bibfnamefont {Q.}~\bibnamefont {Chen}},\ and\ \bibinfo {author}
  {\bibfnamefont {Q.}~\bibnamefont {Liu}},\ }\bibfield  {title} {\bibinfo
  {title} {{Lattice Boltzmann methods for multiphase flow and phase-change heat
  transfer}},\ }\href {https://doi.org/10.1016/j.pecs.2015.10.001} {\bibfield
  {journal} {\bibinfo  {journal} {Progress in Energy and Combustion Science}\
  }\textbf {\bibinfo {volume} {52}},\ \bibinfo {pages} {62} (\bibinfo {year}
  {2016})},\ \Eprint {https://arxiv.org/abs/1508.00940} {arXiv:1508.00940}
  \BibitemShut {NoStop}%
\bibitem [{\citenamefont {Huang}\ \emph {et~al.}(2021)\citenamefont {Huang},
  \citenamefont {Wu},\ and\ \citenamefont {Adams}}]{Huang2021}%
  \BibitemOpen
  \bibfield  {author} {\bibinfo {author} {\bibfnamefont {R.}~\bibnamefont
  {Huang}}, \bibinfo {author} {\bibfnamefont {H.}~\bibnamefont {Wu}},\ and\
  \bibinfo {author} {\bibfnamefont {N.~A.}\ \bibnamefont {Adams}},\ }\bibfield
  {title} {\bibinfo {title} {{Mesoscopic Lattice Boltzmann Modeling of the
  Liquid-Vapor Phase Transition}},\ }\href
  {https://doi.org/10.1103/PhysRevLett.126.244501} {\bibfield  {journal}
  {\bibinfo  {journal} {Physical Review Letters}\ }\textbf {\bibinfo {volume}
  {126}},\ \bibinfo {pages} {244501} (\bibinfo {year} {2021})},\ \Eprint
  {https://arxiv.org/abs/2106.01557} {arXiv:2106.01557} \BibitemShut {NoStop}%
\bibitem [{\citenamefont {Qin}\ \emph {et~al.}(2019{\natexlab{b}})\citenamefont
  {Qin}, \citenamefont {{Del Carro}}, \citenamefont {{Mazloomi Moqaddam}},
  \citenamefont {Kang}, \citenamefont {Brunschwiler}, \citenamefont {Derome},\
  and\ \citenamefont {Carmeliet}}]{Qin2019}%
  \BibitemOpen
  \bibfield  {author} {\bibinfo {author} {\bibfnamefont {F.}~\bibnamefont
  {Qin}}, \bibinfo {author} {\bibfnamefont {L.}~\bibnamefont {{Del Carro}}},
  \bibinfo {author} {\bibfnamefont {A.}~\bibnamefont {{Mazloomi Moqaddam}}},
  \bibinfo {author} {\bibfnamefont {Q.}~\bibnamefont {Kang}}, \bibinfo {author}
  {\bibfnamefont {T.}~\bibnamefont {Brunschwiler}}, \bibinfo {author}
  {\bibfnamefont {D.}~\bibnamefont {Derome}},\ and\ \bibinfo {author}
  {\bibfnamefont {J.}~\bibnamefont {Carmeliet}},\ }\bibfield  {title} {\bibinfo
  {title} {{Study of non-isothermal liquid evaporation in synthetic micro-pore
  structures with hybrid lattice Boltzmann model}},\ }\href
  {https://doi.org/10.1017/jfm.2019.69} {\bibfield  {journal} {\bibinfo
  {journal} {Journal of Fluid Mechanics}\ }\textbf {\bibinfo {volume} {866}},\
  \bibinfo {pages} {33} (\bibinfo {year} {2019}{\natexlab{b}})}\BibitemShut
  {NoStop}%
\bibitem [{\citenamefont {W{\"{o}}hrwag}\ \emph {et~al.}(2018)\citenamefont
  {W{\"{o}}hrwag}, \citenamefont {Semprebon}, \citenamefont {Moqaddam},
  \citenamefont {Karlin},\ and\ \citenamefont {Kusumaatmaja}}]{Wohrwag2018}%
  \BibitemOpen
  \bibfield  {author} {\bibinfo {author} {\bibfnamefont {M.}~\bibnamefont
  {W{\"{o}}hrwag}}, \bibinfo {author} {\bibfnamefont {C.}~\bibnamefont
  {Semprebon}}, \bibinfo {author} {\bibfnamefont {A.~M.}\ \bibnamefont
  {Moqaddam}}, \bibinfo {author} {\bibfnamefont {I.}~\bibnamefont {Karlin}},\
  and\ \bibinfo {author} {\bibfnamefont {H.}~\bibnamefont {Kusumaatmaja}},\
  }\bibfield  {title} {\bibinfo {title} {{Ternary Free-Energy Entropic Lattice
  Boltzmann Model with a High Density Ratio}},\ }\href
  {https://doi.org/10.1103/PhysRevLett.120.234501} {\bibfield  {journal}
  {\bibinfo  {journal} {Physical Review Letters}\ }\textbf {\bibinfo {volume}
  {120}},\ \bibinfo {pages} {234501} (\bibinfo {year} {2018})}\BibitemShut
  {NoStop}%
\bibitem [{\citenamefont {Succi}(2018)}]{Succi2018}%
  \BibitemOpen
  \bibfield  {author} {\bibinfo {author} {\bibfnamefont {S.}~\bibnamefont
  {Succi}},\ }\href {https://doi.org/10.1093/oso/9780199592357.001.0001}
  {\bibinfo {title} {{The Lattice Boltzmann Equation: For Complex States of
  Flowing Matter}}} (\bibinfo {year} {2018})\BibitemShut {NoStop}%
\bibitem [{\citenamefont {Benzi}\ \emph {et~al.}(2009)\citenamefont {Benzi},
  \citenamefont {Chibbaro},\ and\ \citenamefont {Succi}}]{Benzi2009}%
  \BibitemOpen
  \bibfield  {author} {\bibinfo {author} {\bibfnamefont {R.}~\bibnamefont
  {Benzi}}, \bibinfo {author} {\bibfnamefont {S.}~\bibnamefont {Chibbaro}},\
  and\ \bibinfo {author} {\bibfnamefont {S.}~\bibnamefont {Succi}},\ }\bibfield
   {title} {\bibinfo {title} {{Mesoscopic lattice Boltzmann modeling of flowing
  soft systems}},\ }\href {https://doi.org/10.1103/PhysRevLett.102.026002}
  {\bibfield  {journal} {\bibinfo  {journal} {Physical Review Letters}\
  }\textbf {\bibinfo {volume} {102}},\ \bibinfo {pages} {2} (\bibinfo {year}
  {2009})}\BibitemShut {NoStop}%
\bibitem [{\citenamefont {Higuera}\ and\ \citenamefont
  {Succi}(1989)}]{Higuera1989a}%
  \BibitemOpen
  \bibfield  {author} {\bibinfo {author} {\bibfnamefont {F.~J.}\ \bibnamefont
  {Higuera}}\ and\ \bibinfo {author} {\bibfnamefont {S.}~\bibnamefont
  {Succi}},\ }\bibfield  {title} {\bibinfo {title} {{Simulating the flow around
  a circular cylinder with a lattice boltzmann equation}},\ }\href
  {https://doi.org/10.1209/0295-5075/8/6/005} {\bibfield  {journal} {\bibinfo
  {journal} {Epl}\ }\textbf {\bibinfo {volume} {8}},\ \bibinfo {pages} {517}
  (\bibinfo {year} {1989})}\BibitemShut {NoStop}%
\bibitem [{\citenamefont {Stauber}\ \emph {et~al.}(2015)\citenamefont
  {Stauber}, \citenamefont {Wilson}, \citenamefont {Duffy},\ and\ \citenamefont
  {Sefiane}}]{Stauber2015a}%
  \BibitemOpen
  \bibfield  {author} {\bibinfo {author} {\bibfnamefont {J.~M.}\ \bibnamefont
  {Stauber}}, \bibinfo {author} {\bibfnamefont {S.~K.}\ \bibnamefont {Wilson}},
  \bibinfo {author} {\bibfnamefont {B.~R.}\ \bibnamefont {Duffy}},\ and\
  \bibinfo {author} {\bibfnamefont {K.}~\bibnamefont {Sefiane}},\ }\bibfield
  {title} {\bibinfo {title} {{On the lifetimes of evaporating droplets with
  related initial and receding contact angles}},\ }\bibfield  {journal}
  {\bibinfo  {journal} {Physics of Fluids}\ }\textbf {\bibinfo {volume} {27}},\
  \href {https://doi.org/10.1063/1.4935232} {10.1063/1.4935232} (\bibinfo
  {year} {2015})\BibitemShut {NoStop}%
\bibitem [{\citenamefont {Costigliola}\ \emph {et~al.}(2019)\citenamefont
  {Costigliola}, \citenamefont {Heyes}, \citenamefont {Schr{\o}der},\ and\
  \citenamefont {Dyre}}]{Costigliola2019}%
  \BibitemOpen
  \bibfield  {author} {\bibinfo {author} {\bibfnamefont {L.}~\bibnamefont
  {Costigliola}}, \bibinfo {author} {\bibfnamefont {D.~M.}\ \bibnamefont
  {Heyes}}, \bibinfo {author} {\bibfnamefont {T.~B.}\ \bibnamefont
  {Schr{\o}der}},\ and\ \bibinfo {author} {\bibfnamefont {J.~C.}\ \bibnamefont
  {Dyre}},\ }\bibfield  {title} {\bibinfo {title} {{Revisiting the
  Stokes-Einstein relation without a hydrodynamic diameter}},\ }\bibfield
  {journal} {\bibinfo  {journal} {Journal of Chemical Physics}\ }\textbf
  {\bibinfo {volume} {150}},\ \href {https://doi.org/10.1063/1.5080662}
  {10.1063/1.5080662} (\bibinfo {year} {2019})\BibitemShut {NoStop}%
\bibitem [{\citenamefont {Oron}\ \emph {et~al.}(1997)\citenamefont {Oron},
  \citenamefont {Davis},\ and\ \citenamefont {Bankoff}}]{Oron1997}%
  \BibitemOpen
  \bibfield  {author} {\bibinfo {author} {\bibfnamefont {A.}~\bibnamefont
  {Oron}}, \bibinfo {author} {\bibfnamefont {S.~H.}\ \bibnamefont {Davis}},\
  and\ \bibinfo {author} {\bibfnamefont {S.~G.}\ \bibnamefont {Bankoff}},\
  }\bibfield  {title} {\bibinfo {title} {{Long-scale evolution of thin liquid
  films}},\ }\href {https://doi.org/10.1103/revmodphys.69.931} {\bibfield
  {journal} {\bibinfo  {journal} {Reviews of Modern Physics}\ }\textbf
  {\bibinfo {volume} {69}},\ \bibinfo {pages} {931} (\bibinfo {year}
  {1997})}\BibitemShut {NoStop}%
\end{thebibliography}%
\end{document}